\documentstyle[prl,aps,twocolumn,epsfig]{revtex}
\begin{document}

\title{Thermodynamic and electronic properties of a tight-binding lattice-gas
model }

\author{M. Reinaldo-Falag\'an $^1$, P. Tarazona$^1$, E. Chac\'on$^2$, 
J. P. Hernandez$^3$ }

\address{
$^1$Departamento de F\'{\i}sica Te\'orica de la Materia Condensada (C-V),
and Instituto Nicol\'as Cabrera,
Universidad Aut\'onoma de Madrid, E-28049 Madrid, Spain }

\address{
$^2$Instituto de Ciencia de Materiales de Madrid, Consejo Superior de
Investigaciones Cient\'{\i}ficas, E-28049 Madrid, Spain}

\address{
$^3$Department of Physics and Astronomy, University of North Carolina,
Chapel Hill NC 27599-3255 USA }

\date{\today}

\maketitle

Short title: tight-binding lattice-gas model


\vspace{1cm}

{\bf Abstract}

\vspace{1cm}

Thermodynamic and electronic properties are obtained for a lattice-gas
model fluid with self-consistent, partial, occupation of its sites; the 
self consistency consists in obtaining ionic configurations from 
grand-canonical Monte Carlo simulations based on fits to the exact, electronic, 
tight-binding energies of isothermal ensembles of those same ionic 
configurations. The energy of an ion is found to be a concave-up function 
of its local coordination. Liquid-vapor coexistence densities and the 
electrical conductivity, which shows a metal-nonmetal transition, have 
been obtained.

{\bf 1.Introduction}

\vspace{1cm}

    The statistical mechanics of simple insulating liquids is a well 
developed subject, with different approaches being used to obtain the
thermodynamics and the correlation structure, from the additive pairwise 
interaction potential between atoms or molecules \cite{Hansen}. A 
quantum-mechanical study of the atomic or molecular structures provides 
the interatomic potential, but, in all other aspects, the interaction is 
decoupled from the classical statistical-physics problem of obtaining the 
positions and correlations of the atoms.  One of the more striking 
deviations from this {\it simple liquid} behavior is provided by liquid metals,
in which the conduction electrons are fully delocalized and the system 
has to be treated as a mixture of ions ( with classical statistics) and 
electrons (with Fermi-Dirac statistics).  The study of dense liquid 
metals, near the triple-point temperature, has been based on a double 
perturbative expansion \cite{Shimoji} around a {\it reference simple 
fluid} for the ions, and around the {\it jellium} model for the 
conduction electrons. The vapor, at coexistence with the liquid metal, 
has a qualitatively different electronic structure, with the valence 
electrons localized in neutral atoms or clusters. At low temperature, the 
vapor has extremely low density and is almost trivial. In the neighborhood of 
the critical point, and of the metal-nonmetal transition region, the 
interrelation between electronic delocalization and ionic structure 
becomes crucial, and the approaches valid at low temperature fail 
qualitatively.  

Recent experimental data on the critical region of the alkali fluids 
\cite{Gener1} provide strong motivation for a more extensive theoretical 
study of these systems \cite{Stratt,Logan}. Our main objective here is to 
set a minimal model, including the main relevant features of the coupling 
between the electronic and the ionic structure, to analyze the 
statistical-physics problem posed by these systems. We have reported 
\cite{Prelim} preliminary calculations using this approach.  In our 
model, we forsake the 'state of the art' description of liquid metals in 
condensed matter physics, and our model neglects aspects of the problem 
which are certainly relevant for some properties of these systems. 
However, we show that many qualitative effects of the coupling between 
electronic and ionic structures are already present in a very simple 
model. This approach allows the study of the critical region through 
Monte Carlo simulation with system of large size, compared with those 
used in 'state of the art' models, and helps in discerning the crucial 
features which yield a liquid-vapor coexistence curve and an electrical 
conductivity which show that the alkali fluids are qualitatively different 
from the simple fluids.

Even in a very simple model, the exploration of the liquid-vapor 
coexistence in the neighborhood of the critical point requires large 
simulation boxes, which causes multiple exact solutions of the 
tight-binding problem to be rather expensive, computationally. Thus, for 
our Monte Carlo simulation, the electronic free energy will be determined 
using a mapping of the cohesive energies to the local environment of each 
ion, an approach used in our previous work \cite{Tarazona,Chacon} which 
finds stronger justification here. This mapping may be regarded as a 
simplified version of the {\sl glue} model which was proposed and used to 
study the properties of solid and fluid metals \cite{Ercolessi}. The main 
difference with that work is that, in the present case, the mapping is 
not an empirical form used to fit macroscopic parameters (cohesive 
energy, bulk modulus, etc) and then used to obtain other properties of 
the system. Here, the mapping is used to fit exact solutions for an 
ensemble of systems described at microscopic level and the goodness of 
the fit provides a direct check of its quality as trial function, before 
its use to study thermodynamic properties of the system.

\vspace{1cm}

{\bf 2.Tight-binding lattice-gas model}

\vspace{1cm}

Our first simplification consists in using a lattice-gas model to represent 
the ionic structure. Thus, the ions are constrained to partially occupy the 
sites of a chosen lattice. The critical behavior of such 'lattice-gas' 
models, with simple pair interactions, shows that they belong to the same
universality class as simple fluids; so we may expect that the 
long-range fluctuations in the critical region are not qualitatively 
affected by this simplification.  Although in the main body of this work 
the underlying lattice structure is taken to be body centered cubic 
(bcc), since it has the same maximum coordination observed in the 
alkali fluids, we shall also investigate the face centered cubic 
(fcc) to analyze the influence of the lattice on the results.

The electronic properties of our model are described by the simplest 
one-electron tight-binding Hamiltonian for each ionic configuration, with $N$
ions partially occupying the $M$ sites of the lattice. Thus, we postulate 
a single electronic orbital at each occupied lattice site (no orbitals at 
the empty sites) and the electronic hopping is restricted to 
nearest-neighbor (nn) occupied sites. The one-electron energies, 
$\epsilon_n$, and wavefunctions, $\phi_n {(i)}$ (the probability 
amplitudes that an electron occupies the orbital on the ion at site i), 
depend on the ionic configuration. They are the, $n=1, N$, solutions of 
the eigenvalue equation: 
\begin{eqnarray} 
\label{hamil}
(\epsilon_a-\epsilon)  \phi(i)  - t  \sum_j^{nn}  \phi(j) =0,
\end{eqnarray}
for $i=1, N$ and with the sum over $j$ extended to all the occupied sites 
which are nn to site $i$, either directly or through the periodic 
boundary conditions imposed. The site energy, $\epsilon_a$, corresponds to 
that of the atomic orbital and $t$ is the constant hopping parameter between 
occupied nn sites. The one-electron eigenstates are to be occupied, using 
Fermi statistics, with probability 
\begin{eqnarray}
\label{fermi}
f_n=[exp((\epsilon_n-\mu_e)/ k_B T) + 1]^{-1}.
\end{eqnarray}
The electronic chemical potential $\mu_e$ is used to fix the total number 
of electrons to be equal to the total number of atoms. The 
only energy associated with each ionic configuration is the electronic 
free energy: 
\begin{eqnarray}
\label{u}
U = 2 \sum_{n=1}^N \epsilon_n  f_n + 
k_B T [ f_n  ln f_n + (1-f_n) ln (1- f_n) ].  
\end{eqnarray}

If orthonormal solutions to (1) are chosen, charge neutrality at each site
is obtained. An obvious limitation of the present approach is the 
complete neglect of electron-electron interactions. It is noteworthy that 
this neglect results in an overestimate of the single-atom entropy, and an 
underestimate of the experimental dimer energy in comparison with the 
cohesion of the densest system. A similar model for expanded alkali 
metals was proposed by Franz \cite{Franz} using a Cayley-tree 
approximation for the ionic configuration. That work did not study 
the thermodynamic properties of the system, decoupled the ionic and the 
electronic structure, and only calculated the electronic conductivity and 
the magnetic susceptibility.

Structural, thermodynamic, and electronic self consistency demand that
the probability of finding any ionic configuration be proportional to its
Boltzmann factor, $\exp(-(U -N \mu)/k_B T)$, in a grand-canonical ensemble
with chemical potential $\mu$. The self-consistency requirement is 
the same for very rarefied  ($N<<M$) or for very dense ($N \approx M$) 
configurations, although their electronic states are very different. In 
all cases, the preferential double occupancy of the lower energy states 
provides the cohesive energy of the system. Once the underlying lattice 
structure is chosen, the only parameters in the model are $\epsilon_a$ 
and $t$. The parametrization is simplified by taking $\epsilon_a=0$, as 
the energy origin, and using $t$ as the energy scale.

\vspace{1cm}

{\bf 3.Results and discussion}

\vspace{1cm}

Monte Carlo methods can now be used to determine the equilibrium ionic 
structures in conjunction with the exact tight-binding electronic
calculations. We have used simulation cubes with six, seven, and ten bcc 
cells on each side (432, 686, and 2000 sites, respectively) and periodic 
boundary conditions. But, as we have commented previously, this approach is 
computationally expensive and it is therefore interesting to first examine
simple approximations to the problem of finding the statistical equilibrium 
configurations for the ions; these alternate procedures also allow 
obtaining further insight into the problem.

\vspace{0.5cm}

{\bf 3.1.Mean-field approximation}

\vspace{0.5cm}

The simplest approximation which relates the electronic and thermodynamic
properties of our model is a macroscopic mean-field treatment. In this
approximation, electronic feedback to the structure is ignored and the 
ions occupy the lattice sites randomly. It is only required that the lattice 
occupation have a mean density $\rho=N/M$ and that a single ion per 
lattice site be allowed. We then obtain the electronic free energy $U(T)$ in 
(\ref{u}), for a spectrum of realizations of the randomly disordered 
lattice-site occupations which range from nearly-empty to nearly-full 
lattices. An exact diagonalization is carried out for each realization; 
the electronic structure is found to depend on the specific realization. 
Results for the energy per ion, $u=U/N$, as a function of the mean 
density are presented in fig. 1 for the bcc lattice; a typical example with 
$k_B T/t=0.4$ was chosen for this figure. The values of $u$, for different 
realizations with the same value of $\rho$, have a substantial scatter 
around their mean value $\bar{u}(\rho, T)$, particularly at low 
densities. This scatter should scale out with an inverse power of the 
lattice size; in our finite system, the scatter reflects the energies of 
ionic configurations with the same number of ions distributed in 
different (random) patterns. Each isotherm of the average energy per 
particle is a concave-up function. This functional shape is strikingly 
different from the straight-line dependence obtained for nn 
pair-interacting systems. Our lattice model, at this level of 
approximation, is equivalent to a description of the electronic 
properties in continuous models of a fluid, with the correlation 
structure being that of a system of hard spheres \cite{Stratt,Logan,Xu}. 
Such calculations only take into account the packing restrictions, 
without accounting for fluctuations in the clustering. The 
difference between the present approach and those referenced above is 
that now we explore the thermodynamic implications of that electronic 
band structure, with the bold assumption that it provides the only 
cohesive energy.
\begin{figure}
\begin{center}
\epsfig{file=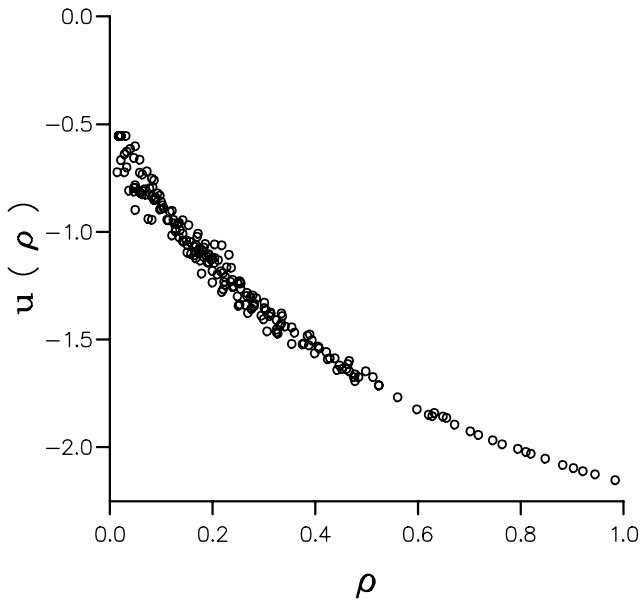,width=8.5cm}
\label{fig1}
\caption[]{The electronic free energy per ion (in units of t)
as a function of the mean fractional-occupation density of the bcc lattice, 
from a random-occupation ensemble and with electronic entropy corresponding 
to $ k_B T/ t = 0.4 $. } 
\end{center}
\end{figure}

In this mean-field approximation, the Helmholtz free energy per ion is 
obtained from $\bar{u}(\rho, T)$ minus $T$ times the ideal lattice-gas 
entropy contribution at each density: 
\begin{eqnarray}
F/N= \bar{u}(\rho, T) + k_B T \left[ \ln(\rho) + {{(1-\rho)} \over \rho}
\ln(1-\rho) \right].
\label{mfa}
\end{eqnarray}
Eq.(\ref{mfa}) can now be used to obtain a liquid-vapor coexistence 
curve; the result is shown as a solid line in fig. 2. The critical 
temperature is found to be $ k_B T_c^{mf}/t=0.56$. The coexistence curve 
obtained exhibits an asymmetry between the densities of the coexisting 
vapor, $\rho_v$, and liquid, $\rho_l$. The asymmetry obtained contrasts 
with the fully symmetric results, $\rho_v=1-\rho_l$, appropriate to a 
lattice gas with nn pair interactions. In that latter case, we know that the 
mean-field approximation overestimates the critical temperature by about 
20\%. In the following sections we shall examine what results arise 
from fluctuations with our tight-binding lattice-gas model.
\begin{figure}
\begin{center}
\epsfig{file=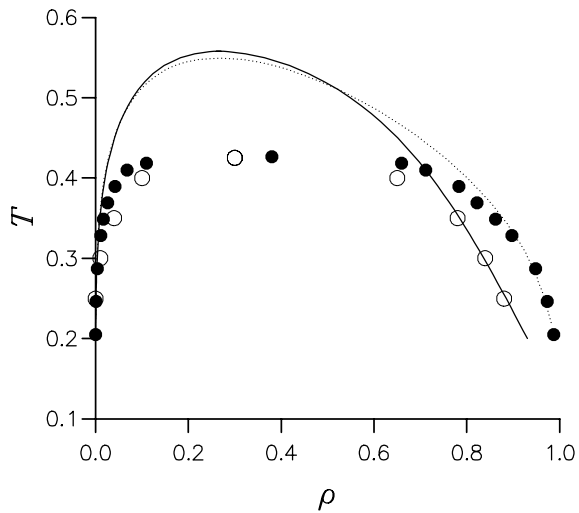,width=8.5cm}
\label{fig2}
\caption[]{Self-consistent results at $ k_B T / t = 0.4 $ for the fitting 
parameters $ u_{k}(T)$, giving the energy (in units of t) of a particle 
with $k$ occupied nn sites, versus $k$ normalized to the lattice 
coordination. Open circles for the bcc lattice with hopping parameter t, 
triangles for the fcc with hopping parameter t, and full circles for fcc 
with hopping parameter $0.82 t$. } 
\end{center}
\end{figure}

\vspace{0.5cm}

{\bf 3.2.Monte Carlo simulation}

\vspace{0.5cm}

In our model, the electronic free energy is determined by first solving
(1) and then occupying the states thermally to obtain the results of (3).
Naturally, the eigenvalue equations depend on the ionic correlations through
the distribution of occupied sites. In turn, as previously mentioned, the
distribution of site occupations is a function of the Boltzmann factors, which
depend on the results of (3). Thus, this is a closed system with the 
distribution of site occupations and the electronic free energy requiring 
self-consistent determination for each set of thermodynamic conditions.

It seems reasonable to attempt a description of the electronic free
energy (3) for each value of $T$, and representative ensembles of 
site occupations, as a function of the nn environments of the occupied 
sites \cite{Tarazona,Chacon}. Thus, we try a least-squares fit, to such 
ensembles of realizations, with a trial form: 
\begin{eqnarray}
\label{env}
U(T) = \sum_{k=0}^{ k_{max} }  u_k(T)  N_k ,
\end{eqnarray} 
where $N_k$ are the number of ions with k occupied nn in each realization
and $ k_{max} $ is the coordination number of the lattice. Ions with $k=0$ 
represent isolated atoms, eq. (3) yields $u_0= - 2 k_B T ln 2$. As 
previously noted, the entropy contribution for isolated atoms is 
overestimated (by a factor of two) in the present tight-binding model. 
The remaining coefficients $u_k$ (for $k=1, k_{max} $) are free, fitting, 
parameters at each temperature. 

For each isotherm, with $k_B T/t$ taking values between 0 and 0.5, we 
begin by using the trial function (\ref{env}) to fit results from the
exact diagonalizations for the random ensemble of realizations used in the
mean-field approximation, at a spectrum of occupation densities. Then, 
using the obtained coefficients $u_k(T)$ as the site energies for 
particles with k occupied nn sites, grand-canonical Monte Carlo 
simulations are performed to obtain equilibrium configurations and 
densities as a function of the chemical potential along the isotherm. 
Then, the process is iterated. A modest sample of equilibrium 
configurations, from those obtained at the chosen temperature and a 
spectrum of the possible system densities, are again exactly 
diagonalized. In contrast to the previous mean-field approach, the ion 
positions are no longer random ones but are correlated to the electronic 
energy as a function of local configuration. The states are occupied 
using the Fermi function and new parameters $u_k(T)$ are obtained from a 
least-squares fit to the new electronic free energies as a function of 
the possible fractional occupation of the lattice. Monte Carlo 
simulations are again performed with the new site energies. The 
procedure is repeated to self consistency at each temperature of 
interest. We then proceed to obtain the self-consistent phase coexistence 
and the electronic properties of the system. The fits, using eq. 
(\ref{env}), are very good, with relative differences always less than 
2\%. An example, obtained as described, of the coefficients $u_k$ for the 
bcc and with $k_B T/t = 0.4$, a near-critical isotherm, is shown as open 
circles in fig. 3. Such results clearly demonstrate the nonlinear 
dependence of the coefficients with the index $k$, expected from the 
nonadditive character of the interactions. A system with nn pair 
interactions would be represented exactly by (\ref{env}) with 
coefficients $u_k$ proportional to the index $k$.
\begin{figure}
\begin{center}
\epsfig{file=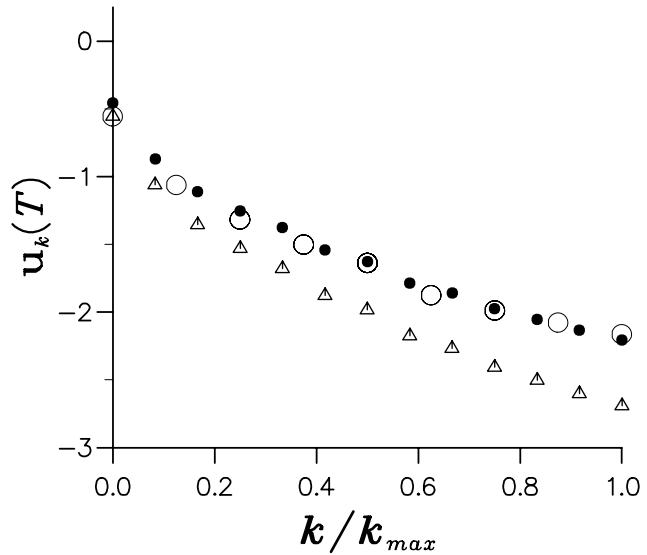,width=8.5cm}
\label{fig3}
\caption[]{Liquid-vapor coexistence curves, with temperature in units of 
$ t/k_B $, vs fractional-occupation density of the lattice, $ \rho$.
The lines are the result of mean-field calculations, eq. (4), with the 
solid one for bcc and the dotted one for fcc (with an electronic hopping 
of $ 0.82 t$). The open circles are the self-consistent bcc Monte Carlo 
results, using the coefficients in eq.(5); similarly, the full circles 
are the fcc results (with an electronic hopping of $0.82 t$). }
\end{center}
\end{figure}

Our simulation boxes are too small to allow a detailed study of the 
critical region. However, the overall features of the self-consistent 
liquid-vapor coexistence curve have been obtained, the results are shown 
as open circles in fig. 2. The critical temperature, in units of the 
hopping parameter, is reduced to about $0.42$ from the previous 
mean-field result of $0.56$. 

We have sought further proof that the number of occupied nn to the 
occupied sites is the important factor in determining the electronic
energy of our system; that is, that the fitting method (eq. (5)) gives
high-accuracy results. As a test, after equilibration, we have generated one 
hundred ionic configurations from successive, complete, Monte Carlo 
sweeps of our lattice, at fixed temperature and chemical potential. The 
exact electronic energy for each of the configurations was obtained and 
compared to that of the fitting method, using the previous set of fixed 
$u_k$ for the chosen temperature. We have chosen a temperature $k_B 
T/t=0.4$ and a chemical potential $\mu/t=-2.275$, as conditions at which 
to perform the grand-canonical Monte Carlo simulations. These conditions 
are near the critical point, where the density has very strong 
fluctuations and the ionic configurations change from isolated clusters 
to complex percolating structures. The fractional-occupation density was 
observed to indeed have large fluctuations, as shown in the upper panel 
of fig. 4 (the line connecting the results for each sweep is shown to 
guide the eye). For each of the ionic configurations obtained, the 
electronic energies per ion are also shown in the lower panel of fig. 4; 
results are exhibited for both the exact diagonalizations of each ionic 
configuration (connected by a solid line) and points arising from the fit 
of eq. (5). The accuracy achieved by the fit of eq. (5) can be seen to be 
impressive, even in this case which has large critical fluctuations.
\begin{figure}
\begin{center}
\epsfig{file=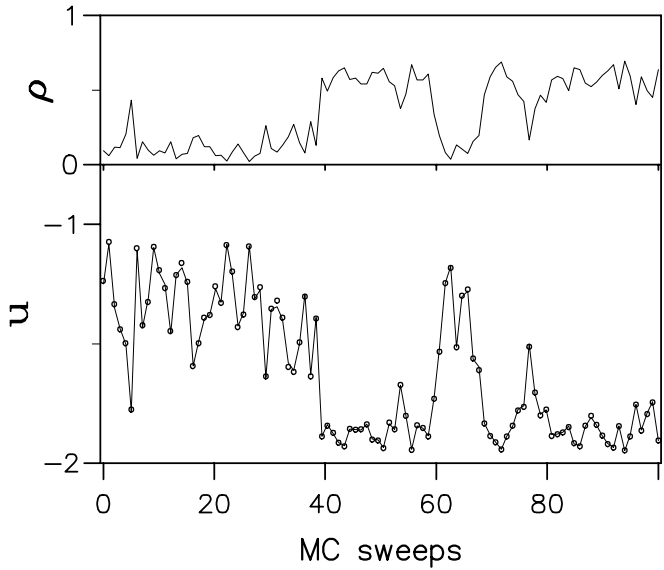,width=8.5cm}
\label{fig4}
\caption{ Fractional-occupation density (upper panel) and electronic free
energy of the system (in units of t, lower panel) for configurations 
resulting, after equilibration, from one hundred, successive, complete, 
Monte Carlo sweeps of the bcc lattice. The thermodynamic conditions are 
close to those appropriate to the critical point: $k_B T/t = 0.4$ and 
$\mu/t = -2.275$. The solid lines (to guide the eye) join the resulting 
densities and also the electronic energies obtained from exact 
diagonalizations of the ionic configurations. The circles give the 
electronic energies from the fit of eq. (5) to the energies of the 
isothermal ensemble, {\it i.e.} from the fixed set $u_{k}(T)$ used in the 
Monte Carlo.}  
\end{center}
\end{figure}

The above results show that the description of the electronic free energy 
in terms of the local coordination of each ion (\ref{env}) is able
to give accurate results for ionic configurations with very different
kinds of disorder. These results support our previous use of this map
\cite{Tarazona,Chacon}, in which the coefficients of (\ref{env}) were
first calculated assuming that the ionic correlation function was that of a
hard-sphere system and then were used to obtain alternatively correlated
structures with strong clustering effects.

\vspace{0.5cm}

{\bf 3.3.Influence of the underlying lattice}

\vspace{0.5cm}

In order to study the influence of the choice of the underlying lattice 
structure on the thermodynamic properties of the model, we have compared 
results obtained using a fcc lattice with the previous ones which were 
based on the bcc. For the fcc system, we have used simulation 
parallelepipeds of five by five by ten cubic cells (1000 sites), and 
periodic boundary conditions. To compare the phase diagrams for a given 
fluid obtained by using different background lattices, we must first 
consider how to identify that the fluid is the same one. We assume that a 
given fluid is specified by its density and cohesive energy, at $T=0$, 
when the lattice is completely full. As the coordination number is 
different for bcc (8) and fcc (12), equating the fluid densities in the 
two systems is easily accomplished by a change in the lattice constant. 
This parameter does not appear explicitly in the calculations, since all 
results are given as depending on the filling fraction. The 
zero-temperature cohesive energy per ion of the completely-full 
lattices are $ u^{fcc} = -2.516 t $  and $ u^{bcc} = -2.064 t $, if the 
same hopping $t$ is used. To make these cohesions equal, with $t$ defining 
hopping in the bcc system, we set $ t^{fcc}= 0.82 t $. At non-zero 
temperatures, all thermal energies will be left scaled to the $t$ for the 
bcc system, while the fcc electronic-energy calculations will take into 
account the modified hopping parameter.

Fig. 3 shows that the fcc parameters $ u_{k} $ (triangles), obtained self 
consistently for the example of a near-critical isotherm, $ k_{B} T /t = 
0.4 $, are quite different from those for the bcc (open circles), if 
scaled to the same $t$. However, when the fcc calculation uses, for the 
electronic energy calculations, a hopping parameter of $0.82 t$ ( full 
circles), the curve almost coincides with the bcc data. Further, in fig. 2 
we also show the fcc liquid-vapor coexistence curves obtained using the 
electronic hopping $ t^{fcc}=0.82 t $. The fcc mean-field approximation 
(dotted line) yields a critical temperature $ k_{B} T_{C}^{mf} / t = 0.55 
$. The liquid-vapor coexistence obtained with the self-consistent map 
(full circles) has a critical temperature $ k_{B} T_{C} / t = 0.42 $. 
Self consistency yields a similar reduction from the mean-field result 
for both lattices. These coexistence curves also exhibit an asymmetry 
between the liquid and vapor branches. When we compare, in fig  2, the 
self-consistent fcc liquid-vapor coexistence curve (with the electronic 
hopping of $ 0.82 t$) with the previous bcc result, both curves can be 
seen to be quite similar; the main discrepancy is in the liquid branch, 
due to the lattice-coordination difference. Given these comparisons, we 
conclude that the choice of background lattice does not strongly affect 
the thermodynamic properties calculated for the fluid under consideration.

\vspace{1.5cm}

{\bf 3.4.Electrical Conductivity}

\vspace{0.5cm}

Our model, although extremely simplified, allows the electrons to couple
with the ionic structure. The effects of this coupling are crucial to the
electronic structure of the model. Mean-field methods with random
occupation of the lattice, like sophisticated treatments in which the
electronic density of states in a liquid is obtained assuming hard-sphere
ionic correlation functions \cite{Stratt}, would give an electronic 
density of states which is independent of the temperature. In  
self-consistent calculations, the electronic structure will reflect the 
dramatic change in the ionic correlations accompanying the thermodynamic 
phase change. This self-consistent interdependence of the electronic and 
ionic structures should dominate the electronic properties of the system. 
We examine some of the consequences of this coupling below, returning to 
the bcc lattice model.

We have explored the relationship between electrical conductivity and ionic 
structure. The experimental signature of a metal-nonmetal transition in
these systems is a decrease of several orders of magnitude in the
conductivity of the expanded fluid. In our previous work 
\cite{Tarazona,Chacon}, we suggested that such behavior is dominantly 
driven by the onset of a lack of percolation of the ionic cluster 
structure leading to a lack of macroscopic delocalization of valence 
electrons, rather than by other features such as a transition to a 
nonmetal due to the Fermi level moving into a regime in which the 
disorder leads to interference-induced (Anderson-type) electronic 
localization. The present model allows probing that assumption, since the 
electronic wavefunctions can be found for any ionic structure.

In our model, it is obvious that the ionic percolation sets a lower bound in
density for the existence of metallic conductivity. We can begin to 
estimate a quantitative measure of the ionic-percolation effects on the 
electrical conductivity thought the Kirchoff's law model proposed by 
Nield {\it et al} \cite{Nield}. Given an ionic structure, this simplest 
approximation replaces each pair of occupied nn sites by a classical, 
fixed-value, resistor. The resistance of the resulting network is calculated 
along the various directions in the simulation box. The value of the 'bond 
resistance' may be fixed to obtain the experimental conductivity in the 
dense liquid (near the triple point) and then the predictions of the 
model, along the coexistence curve or along any isotherm, can be compared 
to the experimental data for the alkali fluids, to see if similar 
features are obtained. 

The above method neglects effects of electronic wavefunctions which may 
preclude electronic conduction even if the ionic structure percolates 
throughout the system. The excellent congruence between the results 
obtained in our previous work \cite{Tarazona,Chacon} and data for fluid 
cesium suggested that electron localization, due to disorder-induced 
interference effects (Anderson-type effects, in contrast to a lack of 
percolation), was not important in the conductivity of expanded alkali 
metals. In the present model, we may compare the results of the above 
simplest Nield model with a quantum estimate for the conductivity; for 
example, the results of assuming a loss of phase memory (due to 
scattering) after a nn hop. This alternative is a variant of the 
randomized phase model of Hindley \cite{Hindley} in the Kubo-Greenwood 
formula \cite{??}, modified for the disordered topology of hopping sites 
in our model. In this alternative approach, the conductance between 
occupied nn sites i and j is given by 
\begin{eqnarray}
1/ R_{i, j} \propto \sum_{m \neq n} 
\frac{\partial f_n}{\partial \epsilon_n}
 |\phi_n (i)|^2 |\phi_m (j)|^2 \delta(\epsilon_n - \epsilon_m),
\end{eqnarray}
which requires a non-zero product, at nn sites, of different electronic
states, $m$ and $n$, which have the same energy, for this elastic
scattering case. In our finite size system, the condition of elastic
scattering will be relaxed, with the delta function being replaced by a 
Gaussian of variance equal to $0.16 t$. Actually, the precise 
Gaussian-width value has little importance to the results, except near 
complete filling of the lattice. The required energies and amplitudes are 
obtained from the exact diagonalizations for typical equilibrium 
configurations at the thermodynamic parameters of interest. Results 
for the macroscopic conductivity are then obtained as in the Nield model 
but using (6) for each 'bond resistor'. This approach partially takes 
into account electron-wavefunction effects on the conductivity. We have 
used it, with phase-memory-loss scattering after a nn hop, to caricature 
the strong scattering to be expected in a hot disordered fluid.

For some particular ionic configurations, near the obtained percolation
threshold, the results of the Kubo-Greenwood calculation could show a
strong dispersion in the conductances of the different bonds due to 
interference effects, which would then be reflected in an overall 
resistance very different from that due to setting all 'bond resistors' 
equal to their mean value, as is done in the simplest approach. For 
example, Franz \cite{Franz}, using a Cayley-tree approximation for the ionic 
configuration and our Kubo-Greenwood formula for the conductivity, 
obtains a critical density for quantum percolation which is higher than for 
classical percolation. However, in the statistical sampling of ionic 
configurations from our model, it was found that most configurations do 
give global resistances which are well described by the simplest version 
of the Kirkchoff's law model. An example of the result of each of our
two procedures is shown in fig. 5, for a near-critical isotherm of the 
bcc lattice. As can be seen, there is good agreement between the 
alternate methods, which indirectly supports our previous hypothesis that 
the nonmetal-metal transition in the alkali fluids is mainly driven by 
the onset of percolation of ionic structures.
\begin{figure}
\begin{center}
\epsfig{file=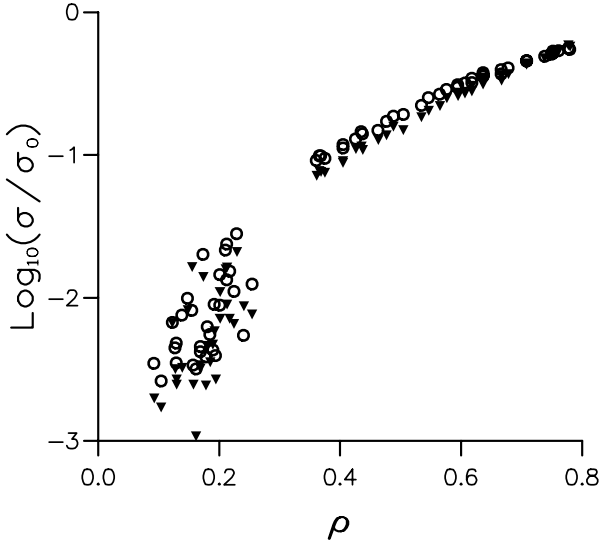,width=8.5cm}
\label{fig5}
\caption[]{Electrical conductivity estimates $\sigma$ for the bcc lattice, 
normalized to the conductivity at full lattice occupation $\sigma_{0}$, 
logarithmic scale vs density, for the isotherm $k_B T/t = 0.4$.  The full 
triangles are obtained using the Greenwood-Kubo quantum values of the 
bond resistors (eq. (6)); the open circles are the results of using a 
fixed bond resistor; both approaches then use the Kirchoff model for a 
spectrum of Monte Carlo configurations.}  
\end{center}
\end{figure}

\vspace{1cm}

{\bf 4.Summary}

\vspace{1cm}

We have presented a tight-binding lattice-gas model, 
which can be solved for its structural and 
thermodynamic properties. The model takes into account the inhomogeneity 
of statistical configurations of the system, driven by nonadditive 
interactions due to valence-electron delocalization. A self-consistent 
procedure is used to determine the equilibrium structures: Monte Carlo 
simulations coupled to exact diagonalizations and site-energy fittings. 

As previously noted, the first qualitative consequence of the feedback from
the electronic to the ionic structures in our model is the existence of a
vapor-liquid phase transition.  At low temperatures, random distributions
of the ions, assumed in studies with frozen disorder, are 
thermodynamically unstable. The instability can cause a condensation into 
drops of dense liquid at coexistence with a vapor at lower density. The 
qualitative features of the condensation which results from our model, 
driven by the electronic band energy, will be compared with experimental 
alkali-fluid data.  Although the extreme simplicity of the present model 
should preclude expectations of quantitative agreement, the phase 
diagrams in fig. 6, our self-consistent results (full and open circles, 
for bcc and fcc respectively) and experimental data for Cs (solid line), 
all in units of the critical temperatures and densities, show an appealing 
similarity. The strong asymmetry, observed in the experiments, between 
the coexisting vapor and liquid densities is also quite clear in the 
results from our model; we emphasize this feature by showing the 
'diameter function' from the fcc calculation. This asymmetry contrasts 
with the coexistence curve results, also shown in fig. 6 and due to 
particle-hole symmetry, of lattice-gas models with additive nn pair 
interactions.  
\begin{figure}
\begin{center}
\epsfig{file=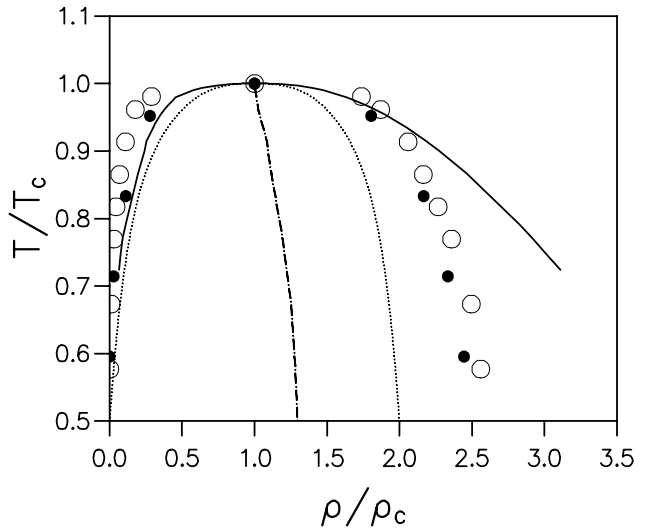,width=8.5cm}
\label{fig6}
\caption[]{Reduced liquid-vapor coexistence curves, temperatures and
densities scaled to their critical values. The full line is a fit to
experimental data for Cs. The circles are the self-consistent 
Monte Carlo results from the present model; full circles for bcc, open 
circles for fcc. The dashed-dotted line is the diameter function $\rho_d 
= ( \rho_L + \rho_V) / 2 $ for the self-consistent Monte Carlo results 
with the fcc lattice. The dotted line is the result for a bcc 
lattice-gas model with additive nn pair interactions.} 
\end{center}
\end{figure}

Our second qualitative result is that, in this model, equilibrium density 
fluctuations only weakly influence the phase diagram. The critical 
temperature in the mean-field approximation is about 20\% higher than the 
value obtained in the Monte Carlo simulation. This difference is similar to 
that obtained for usual models of simple fluids, with pair interactions
in three dimensions. However, such a modest difference contrasts with the 
much larger effect, a ratio of almost a factor of three, in our previous 
work using pseudopotentials \cite{Tarazona,Chacon}. The origin of this 
contrast is that in the previous work both the ion-ion and the ion-electron 
correlations are neglected in the mean-field perturbative treatment using the 
homogeneous electron gas as a reference. The poor description of 
ion-electron correlations, even after the perturbation of the reference 
system, is the principal cause of the high critical temperature which 
results. In contrast, in the present tight-binding model ion-electron 
correlations are treated exactly, there are no direct ion-ion interaction 
to be corrected by the correlations of the  ionic structure, and the 
difference between the mean-field treatment (a random filling of the 
lattice sites) and the simulation is only due to corrections of the 
electronic energy due to the effects of ion-ion correlations.

The electrical conductivity of the system was estimated, using a 
quantum-resistor network and the self-consistent ionic structures. The 
metal-nonmetal transition was then probed in a manner which is unified with
the previous structural treatment. It appears that the main reason the 
material becomes nonmetallic, on lowering its density, is the lack of 
percolation of its self-consistent ionic structure, rather than due to 
more complicated interference-induced electron localization effects.

The present model has provided confirmation that an energy mapping which
reflects the local coordination of each particle gives a good description 
of the system. Thus, we have verified that the approach followed in our 
previous work \cite{Tarazona,Chacon} is justified and adequate. Our 
conclusion that the system becomes metallic depending mainly on the 
self-consistent percolation of the site occupation is based on the 
one-electron approximation. Since we assumed a paramagnetic system and 
did not include Hubbard-type interactions, we cannot speculate on the 
effects of Mott-Hubbard interactions on the metal-nonmetal transition.   
Such effects will be investigated in future work. 

The model is a very simplified representation of a metal-atom fluid, as
only the electronic band energy contributes to the system cohesion.
Nevertheless, the results obtained show that the model contains the basic
ingredients to allow qualitative reproduction of the peculiar behavior
observed in the alkali fluids. These peculiarities include the
metal-nonmetal and liquid-vapor transitions, the general shape of the
coexistence curve, and the connection between ionic and electronic
structures. The self consistency between the electronic and ionic
structures is the feature which allows the model to give a unified
treatment of a wide spectrum of the system properties and to
qualitatively describe observations in the alkali fluids.

\acknowledgements

This work was partially supported by the Direcci\'on General
de Investigaci\'on Cient\'{\i}fica y T\'ecnica (Spain) under Grant
PB94-005-C02 and by the NATO Office of Scientific Research via grant
SA.5-2-05(CRG.940240). One of us (JPH) is also grateful for financial
support to the W. R. Kenan, Jr. Foundation, the Spanish Ministry of Science
and Education, and the Instituto Nicol\'as Cabrera.

\end{document}